\def\to{\hbox{$\,$--$\,$}}
\def\muspc{\hskip 0.15 em}
\def\mmuspc{\hskip 0.05 em}
\def\minus{\hbox{$-$\mmuspc}}
\input psfig
\documentstyle{mnb}
\title{On the galactic disc age-metallicity relation}
\author[G. Carraro et al.]{Giovanni Carraro$^{1,2}$, 
Yuen Keong Ng$^3$ and Laura Portinari$^{1,4}$\\
$^1$ Department of Astronomy, Vicolo dell'Osservatorio 5,
     I-35122 Padova, Italy \\
$^2$ SISSA-ISAS, Via Beirut 2-4, I-34014 Trieste, Italy \\ 
$^3$ Osservatorio Astronomico di Padova, Vicolo dell'Osservatorio 5,
     I-35122 Padova, Italy \\
$^4$ Nordita, Blegdamsvej 17, DK-2100 Copenhagen \OE, Denmark \\
$^{\null}\ $\ E-mail: 
{\tt carraro,yuen,portinari\char64astrpd.pd.astro.it}
}
\date{\it submitted}
\pagerange{\pageref{firstpage}--\pageref{lastpage}}
\pubyear{1997}
\begin{document}
\maketitle

\label{firstpage}

\hyphenation{me-tal-li-city}
\hyphenation{Solar Wielen}
\begin{abstract}
A comparison is made between the age-metallicity relation obtained from four
different types of studies: F and G stars in the Solar Neighbourhood, analysis
of open clusters, galactic structure studies with the stellar population
synthesis technique, and chemical evolution models. Metallicities of open
clusters are corrected for the effects of the radial gradient, which we find
to be \minus0.09~dex~kpc$^{-1}$ and to be most likely constant in time. 
We do not
correct for the vertical gradient, since its existence and value are not firmly
established. 
\par
Stars and clusters trace a similar age-metallicity relation, showing an excess
of rather metal-rich objects in the  age range 5\to9~Gyr. 
Galactic structure studies tend to give a
more metal-poor relation than chemical evolution models.
Both relations do not explain the presence of old, relatively metal-rich stars
and clusters. This might be due to uncertainties in the ages of the local
stars, or to pre-enrichment of the disc with material from the
bulge, possibly as a result of a merger event in the early phases of the
formation of our Galaxy. 
\end{abstract}


\section{Introduction} 
The age-metallicity relation (AMR) for nearby stars is a record of the
progressive chemical enrichment of the star-forming local interstellar medium
during the evolution of the galactic disc, and so provides useful clues 
about the
star formation and chemical evolution history of the local environment.
Metallicity is usually identified with the [Fe/H] ratio. Oxygen would actually
be a better tracer and ``chronometer'' of metal enrichment (Wheeler et~al.\
1989), because it is the most abundant metal and it is produced on the
well-defined and short time-scale of type II SN\ae. On the contrary, iron is
released by various sources (both type II and type Ia SN\ae) with rather
different time-scales. 
Anyway, metallicity is generally determined by the [Fe/H]
ratio, since the [O/H] ratio is much more difficult to measure in stellar
atmospheres. 
\par
In the Solar Neighbourhood the metallicity of the stars can be studied in high
detail (Edvardsson et~al.\ 1993, hereafter Edv93ea), 
but the results are liable to large
errors in the individual determinations of the age.
Ng\mbox{\muspc\&\muspc}Bertelli (1998, hereafter NB98) 
demonstrated that the errors in the age
are mainly due to the large uncertainties in the individual distances.
Distances need to be known to a 5\% accuracy to get reliable ages. In
this respect, the ages and metallicities of clusters are more reliable, since
one is dealing with a group of stars and therefore the result is less
susceptible to individual errors. 
\par
An AMR can also be obtained from star counts studies, based on the population
synthesis technique (Bertelli et~al.\ 1995, 1996; Ng et~al.\ 1995, 1996, 1997).
In such studies all the stars along the line of sight are considered. The disc
is sampled with respect to age and metallicity 
in layers with specific effective
thicknesses. In this way indications are obtained for the disc's chemical
evolution. 
\par
The aim of this paper is to compare the AMR obtained from various methods and
to critically discuss the probable causes for the differences found. In Sect.~2
we start with a general overview of the various AMRs and improve them when
possible. In Sect.~3 the relations are compared with each other and probable
causes for any discrepancy are outlined. The results are finally summarized in
Sect.~4. 

\section{Age-Metallicity relation}
In this section we describe the AMRs considered for this study. We start with
the AMR obtained from stars in the Solar Neighbourhood. Then we consider the
AMR obtained from open clusters, by improving the results from
Carraro\mbox{\muspc\&\muspc}Chiosi (1994a, hereafter CC94a) 
with a larger cluster sample.
We continue with the description of the AMR obtained
from star counts analysis based on the stellar population synthesis technique.
Finally, we discuss the AMR obtained from chemical evolution models. 
\begin{figure}
\centerline{\vbox{\psfig{file=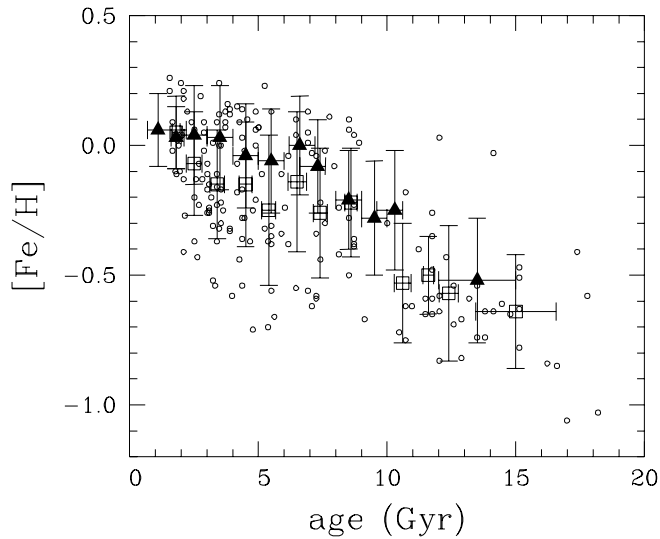,height=6.91cm,width=8.5cm}}}
\caption{AMR for nearby stars by Meusinger et~al.\ (1991; triangles) and
Edvardsson et~al.\ (1993, open squares and open circles).}
\label{AMRedv}
\end{figure}

\subsection{The Solar Neighbourhood}
\label{SolarN}
Twarog (1980) first showed that nearby stars display 
an AMR, by applying {\it
uvby-$\beta$} photometry and Yale isochrones to get the metallicity and the age
for two wide samples of stars. Since single data showed a remarkable
dispersion, age-bins and average metallicity per bin were used to deduce the
local AMR. The metallicity turned out to increase sensitively during disc
evolution, rapidly from 13 to 5 Gyr ago, and more slowly onwards. 
\par
Twarog's data were later re-examined by means of different photometric
calibrations with the Vandenberg (1983, 1985) isochrones, resulting in
discordant AMRs.
\hfill\break
Carlberg et~al.\ (1985) limited themselves to only one of the two samples from
Twarog, hereby excluding around fifty of the lowest metallicity stars. They
obtained a relatively high average metallicity for old stars and a shallower
slope of the AMR in the early phases, suggesting that the disc evolution
started with a high initial metallicity. 
\hfill\break
On the other hand, Meusinger et~al.\ (1991) re-examined both samples from
Twarog with Vandenberg's isochrones and with a new calibration for the
metallicity index and improved model atmospheres. The resulting AMR was closer
to the one obtained by Twarog and had a considerable slope toward old ages.
This relation is displayed in Fig.~\ref{AMRedv}. 
\par
\begin{figure}
\centerline{\vbox{\psfig{file=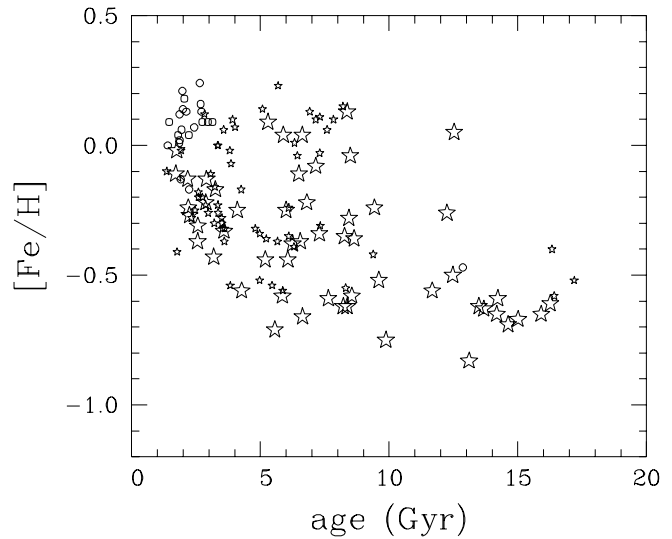,height=6.91cm,width=8.5cm}}}
\caption{The AMR for nearby stars by 
Ng\mbox{\muspc\&\muspc}Bertelli (1997). 
The ages are computed with the distances obtained from 
the {\it Hipparcos}\/ parallax (ESA 1997).
The open circles indicate star on
or near the main sequence with an uncertainty in the age less than 12\%.
Open stars are used for giant branch stars. The uncertainty
in the age is less than 12\% for the large symbols and greater than 
12\% for the small symbols.}
\label{AMRng}
\end{figure}
A new estimate of the local AMR, based on a sample of 189 nearby F and G
dwarfs, was performed by Edv93ea. The metallicity is derived
from the analysis of high resolution spectra with theoretical LTE model
atmospheres. The ages were determined from fits of the photometric data with
Vandenberg (1985) isochrones. The resulting binned AMR is in good agreement,
within the large dispersion, with that by Meusinger et~al.
Figure~\ref{AMRedv} shows the averaged, binned AMR (open squares) from
Edv93ea together with the individual data points. 
\hfill\break
The Edv93ea dataset was re-analysed by
NB98 by means of Bertelli et~al.\ (1994)
isochrones, based on the latest opacity tables. 
Near the main sequence, high age isochrones 
are packed closely together and the deduced ages have
considerable uncertainties. The situation changes if a star is on the giant
branch: its evolution is relatively fast and 
the uncertainty in the age becomes
considerably smaller. For the determination of the ages,
NB98 also took into account the uncertainties
on the effective temperature, metallicity and distance of each star; the
distances of the stars are obtained from the {\it Hipparcos}\/ (ESA 1997)
parallaxes. In this way, reliable ages were obtained for the stars displayed 
with big symbols in
Fig.~\ref{AMRng}. The resulting AMR has a small, but
distinct slope of \mbox{$\sim$\muspc0.07 dex/Gyr} when the stars older than
10~Gyr are considered. However, NB98 suspect that
the ages for $t\!>\!10$~Gyr might be overestimated ($\sim$\muspc2~Gyr), due to
an improper selection of the isochrones for these stars, 
which are over-abundant
in $\alpha$-elements. It is noteworthy that this relation is essentially not
different from the relation one gets when, for the stars with reliable ages,
the ages from Edv93ea are adopted; we refer to
NB98 for additional details. 
\par
However, the most striking feature of the results mentioned above is the huge
scatter around the average trend, which makes the correlation between age and
metallicity rather weak, especially for stars with $t\!<\!10$~Gyr. Such a
spread in the data must be in part intrinsic, and many possible causes
have been suggested: orbital diffusion of stars coupled with radial metallicity
gradients, local inhomogeneities in the star-forming gas, inhomogeneous
accretion of infalling external gas, overlapping of different galactic
sub-structures, each with its own specific AMR. All the mentioned effects may
actually contribute to produce the observed scatter, and the new challenge for
chemical models is nowadays to explain such a large spread in the local AMR
(see Sect.~\ref{chemmodels}). 

\begin{table}
\tabcolsep 0.10truecm
\caption{Sample of 37 clusters with: ages determined by the synthetic CMDs
technique (2$^{nd}$ column), spectroscopic metallicities in the
Boston scale (Friel\mbox{\muspc\&\muspc}Janes 1993, Friel 1995; 3$^{rd}$ 
column) and corresponding errors (4$^{th}$ column), height above the galactic
plane (5$^{th}$ column) and galactocentric distance (6$^{th}$ column).}
\begin{center}
\begin{tabular}{lcccrc}\hline
\multicolumn{1}{c}{Cluster}&
\multicolumn{1}{c}{age(Gyr)}&
\multicolumn{1}{c}{\phantom{+}[Fe/H]}&
\multicolumn{1}{c}{r.m.s.}&
\multicolumn{1}{c}{$z$ (pc)}&
\multicolumn{1}{c}{R (kpc)$^\dagger$}\\
\hline
Berkeley 17&  9.00&\phantom{+}\minus0.29&  0.13&   170&  11.2\\
Berkeley 19&  3.80&\phantom{+}\minus0.50&  0.10&   300&  13.3\\ 
Berkeley 20&  3.00&\phantom{+}\minus0.75&  0.21&  2100&  16.0\\
Berkeley 21&  3.10&\phantom{+}\minus0.97&  0.10&   260&  14.5\\ 
Berkeley 31&  4.00&\phantom{+}\minus0.50&  0.16&   430&  12.7\\
Berkeley 32&  3.00&\phantom{+}\minus0.58&  0.10&   300&  12.0\\
Berkeley 39&  6.50&\phantom{+}\minus0.31&  0.08&   705&  11.7\\
Cr 261&       7.00&\phantom{+}\minus0.14&  0.14&   420&\phantom{1}7.6\\
IC 166&       0.85&\phantom{+}\minus0.32&  0.20&    10&  10.7\\ 
IC 4651&      1.60&\phantom{+}\minus0.16&  0.05&   100&\phantom{1}7.8\\ 
King 5&       0.80&\phantom{+}\minus0.38&  0.20&   180&  10.5\\
King 11&      6.00&\phantom{+}\minus0.36&  0.14&   460&  10.5\\
M 67&         4.80&\phantom{+}\minus0.09&  0.07&   415&\phantom{1}9.1\\ 
Melotte 66&   5.46&\phantom{+}\minus0.51&  0.11&   710&\phantom{1}9.4\\ 
NGC 188&      7.50&\phantom{+}\minus0.05&  0.11&   570&\phantom{1}9.3\\ 
NGC 752&      1.50&\phantom{+}\minus0.16&  0.05&   145&\phantom{1}8.7\\ 
NGC 1193&     5.00&\phantom{+}\minus0.50&  0.18&  1020&  12.7\\ 
NGC 1245&     0.80&\phantom{\minus}+0.14&  0.10&   460&  11.1\\
NGC 1817&     0.80&\phantom{+}\minus0.39&  0.04&   410&  10.3\\ 
NGC 2141&     2.50&\phantom{+}\minus0.39&  0.11&   430&  12.6\\  
NGC 2158&     1.42&\phantom{+}\minus0.23&  0.07&    95&  11.6\\ 
NGC 2204&     1.74&\phantom{+}\minus0.58&  0.10&  1200&  11.8\\ 
NGC 2243&     4.50&\phantom{+}\minus0.56&  0.17&  1260&  11.1\\ 
NGC 2420&     2.10&\phantom{+}\minus0.42&  0.07&   655&  10.3\\ 
NGC 2477&     0.60&\phantom{+}\minus0.05&  0.11&   135&\phantom{1}9.0\\
NGC 2506&     1.90&\phantom{+}\minus0.52&  0.07&   460&  10.4\\ 
NGC 2660&     0.70&\phantom{\minus}+0.06&  0.10&   152&\phantom{1}9.2\\ 
NGC 3680&     1.80&\phantom{+}\minus0.16&  0.05&   230&\phantom{1}8.3\\ 
NGC 3960&     0.60&\phantom{+}\minus0.34&  0.08&   180&\phantom{1}8.0\\
NGC 5822&     0.45&\phantom{+}\minus0.21&  0.12&    45&\phantom{1}7.9\\
NGC 6791&     8.00&\phantom{\minus}+0.19&  0.19&  1010&\phantom{1}8.4\\ 
NGC 6819&     2.05&\phantom{\minus}+0.05&  0.11&   310&\phantom{1}8.2\\ 
NGC 6939&     1.40&\phantom{+}\minus0.11&  0.10&   255&\phantom{1}8.7\\ 
NGC 6940&     0.60&\phantom{\minus}+0.04&  0.10&   100&\phantom{1}8.3\\ 
NGC 7142&     4.90&\phantom{+\minus}0.00&  0.06&   495&\phantom{1}9.7\\ 
NGC 7789&     1.35&\phantom{+}\minus0.26&  0.06&   165&\phantom{1}9.4\\ 
Tombaugh 2&   1.75&\phantom{+}\minus0.70&  0.18&  1030&  15.6\\ 
\hline
\multicolumn{6}{c}{\footnotesize $^\dagger$\quad R is computed with
$R_{\odot}\!=\!8.5$~kpc 
for the distance of the}\\
\multicolumn{6}{c}{\footnotesize Sun to the galactic centre 
for consistency with previous work.}\\ 
\end{tabular}
\end{center}
\end{table}

\subsection{Open clusters}
\label{openclusters}

\subsubsection{Sample}
Most of the open clusters considered in this paper (Table~1) are taken from the 
compilation presented by CC94a. We selected from that sample all the clusters
for which a homogeneous metallicity determination was available:
all the metallicities in Table~1 are in fact obtained spectroscopically 
and they are in the Boston scale (Friel\mbox{\muspc\&\muspc}Janes 1993,
Friel 1995). Therefore, with respect to CC94a
we do not consider here NGC~6603, IC~1311, NGC~7704, King~2 and AM~2, since
for these cluster the lack of a homogeneous metallicity estimate prevents 
us from obtaining their age with the same method as for all the other clusters.
\hfill\break
In addition, the CC94a sample has been updated including here
Berkeley~17 (Kaluzny 1994), Berkeley~20, Berkeley~31 and King~5 (Phelps et~al.\
1994), Collinder~261 (Mazur et~al.\ 1995), and NGC~1245 
(Carraro\mbox{\muspc\&\muspc}Patat 1994). Table~1 lists our updated 
compilation for 37 clusters.
\par
In CC94a, the ages of ten clusters were determined with stellar 
isochrones using
the synthetic Colour--magnitude diagrams  (CMDs) technique (Chiosi et~al.\
1989). This sample defined a relation between age and 
$\Delta$V (cf.~Cannon 1970, Cameron 1985, 
Barbaro\mbox{\muspc\&\muspc}Pigatto 1984,
Anthony-Twarog\mbox{\muspc\&\muspc}Twarog 1985).
The ages of the remaining clusters were obtained through interpolation
of this relation, only in a few cases an extrapolation
was applied.
von Hippel et~al.\ (1995) noticed for one cluster a discrepancy between the
extrapolated $\Delta$V age with the age determined from the white dwarf
cooling sequence. This however does not imply, as concluded by
von Hippel et~al.\ (1995), that the isochrone ages are in error.
It mainly demonstrates that one has to be cautious with extrapolations.

In this paper, the ages for all the clusters in Table~1
have been determined by fits with the Bertelli et~al.\ 
(1994) isochrones using the synthetic CMDs technique, 
following the procedure described by
Carraro\mbox{\muspc\&\muspc}Chiosi (1994a, 1995).
We found for Berkeley~17 and Collinder~261
an age sensitively different
from recent studies (Phelps~1997 and Gozzoli et~al.\
1996 respectively). In the case of Berkeley~17 the different value can 
be ascribed to the use of different datasets and different techniques
(simple isochrones fitting against synthetic CMDs). 
The same partly holds for Collinder~261. We used a different
dataset and  a slightly different method. Gozzoli et~al.\ 1996 
derived the age from evolutionary tracks 
(Fagotto et~al.\ 1994), while we determined the age from
suitably interpolated isochrones (Bertelli et~al.\ 1994).
Finally, we performed simulations with synthetic CMDs and
used the capability of our code to interpolate 
between isochrones of different metallicities,
adopting for the simulations 
a metallicity $Z$ corresponding to the observational [Fe/H].
\par
The main advantage of this compilation is the homogeneity
of the sample: ages, metallicities and positions in 
the galactic disc are all obtained in the same fashion.
This homogeneity is not guaranteed when using larger samples
as in Friel (1995), Piatti et~al.\ (1995, hereafter P95ea)
or Twarog~et~al.\ (1997).
In addition, completeness is a crucial ingredient 
in order to explore possible relations between the 
cluster properties and their position in the galactic disc.
However, it is not possible to gather a complete sample for
the old, open clusters in the galactic disc, because of strong selection 
effects mainly related to the past dynamical history of the galactic disc.
Many old clusters have probably been disrupted (van den Bergh \& McClure
1980; Friel 1995).
Taken this limitation into account, 
we used a statistical approach 
to find correlations between the clusters'
fundamental parameters, i.e. age, metal abundance and position
inside the galactic disc.

\begin{figure*}
\centerline{\vbox{\psfig{file=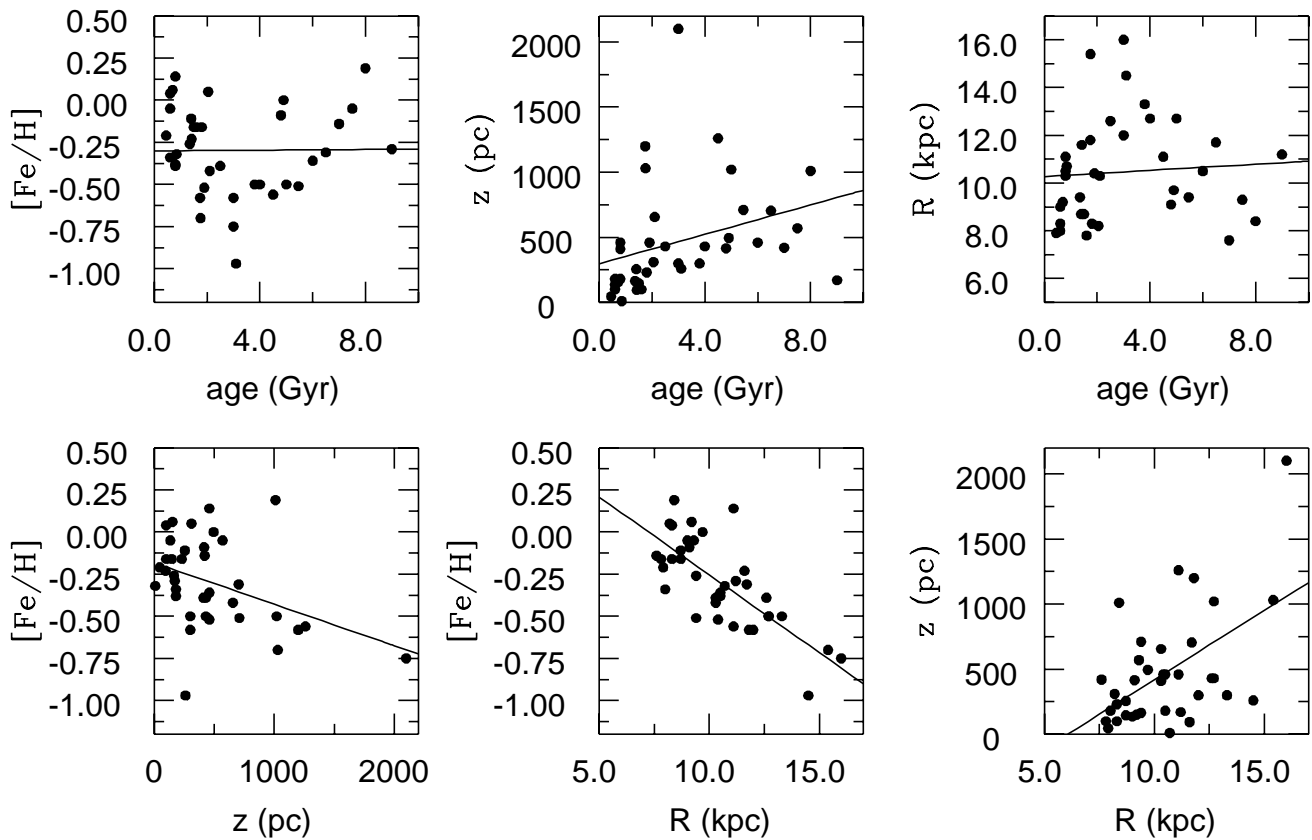,height=11.0cm,width=17.5cm}}}
\caption{Relation between the fundamental parameters of the open 
cluster sample given in Table~1,
see Sects.~2.2.2\mbox{\muspc\&\muspc}2.2.3 for details.
Solid lines are the projections of the first principal component:
([Fe/H]\to{$t_9$}), ($z$\to{$t_9$}), (R\to{$t_9$}),
([Fe/H]\to{$z$}), ([Fe/H]\to{R}), and ($z$\to{R}).}
\label{pca}
\end{figure*}

\subsubsection{Multivariate analysis}
\label{mva}
The search for correlations has been realized by means of Multivariate Data
Analysis (Murtagh\mbox{\muspc\&\muspc}Heck 1987) 
with the SPSS package (Nie et~al.\ 1975).
In particular, we performed the so-called Principal Component Analysis
(PCA), which is designed to find the number of independent parameters
characterizing multivariate data.
This technique is widely used in astronomy; for applications to other
astronomical problems we refer to Murtagh\mbox{\muspc\&\muspc}Heck.
\par
The crux of the method is to search for suitable linear coordinate
transformations, which are by definition orthogonal,
and to diagonalize the covariance matrix from an 
original set of variables ($x_{1}, x_{2},...x_{n}$)
to a new set of variables 
($\psi_{1},\psi_{2},...\psi_{n}$)
with a zero mean value:

\begin{equation}
\quad\psi_{i} = \sum_{k=1}^{n} l_{ik} x_{k} ,
\end{equation}

\noindent
where $l_{ik}$ are the coefficients of the transformation.
The properties of the transformation are such, that the eigenvalues 
$\lambda_{i}$ of the matrix $<\!\psi_{i} \psi_{j}\!>$ 
are the variances of the data in the direction
of the principal component.
The corresponding eigenvectors are then used to build
the new linear combinations from the initial parameters.
By convention, the first principal component corresponds to the largest
eigenvalue. As a consequence, the first principal component is a minimum
distance fit to a line in the space of the original parameters.
The same holds for a possible second, third and so on principal component.
\par
\begin{table}
\tabcolsep 0.10truecm
\caption{Principal components ($\psi_{1},\psi_{2}$) 
and eigenvalues for 
the open cluster sample listed in table~1,
together with the values ($\psi_{1p},\psi_{2p}$) 
for the Piatti et~al.\ (1995) sample.}
\begin{center}
\begin{tabular}{lcccccc}
\hline
\multicolumn{1}{c}{$l_{ij}$}&
\multicolumn{1}{c}{age}&
\multicolumn{1}{c}{[Fe/H]}&
\multicolumn{1}{c}{$z$}&
\multicolumn{1}{c}{R}&
\multicolumn{1}{c}{$\lambda$}&
\multicolumn{1}{c}{$variance(\%)$}\\
\hline
$\psi_{1}$&0.201&\minus0.849&\phantom{\minus}0.755&\phantom{\minus}0.914&2.167&54.2\\
$\psi_{2}$&0.938&\phantom{\minus}0.284&\phantom{\minus}0.256&\minus0.153&1.049&26.2\\
\\
$\psi_{1p}$&0.624&\minus0.890&\minus0.331&\phantom{\minus}0.782&1.903&47.6\\
$\psi_{2p}$&0.619&\phantom{\minus}0.100&\phantom{\minus}0.873&\minus0.011&1.155&28.9\\
\hline
\end{tabular}
\end{center}
\end{table}

\subsubsection{Principal Components}
\label{pcs}
In our case the original space of parameters is four-dimensional. 
The parameters are the age, the metallicity, the $z$-coordinate and
the radial distance R from the galactic centre 
(R is computed with
$R_{\odot}\!=\!8.5$~kpc, where $R_{\odot}$
is the distance of the Sun to the galactic centre).
The straight application of PCA leads to the identification of two 
principal components shown in Table~2. 
\begin{figure*}
\centerline{\vbox{\psfig{file=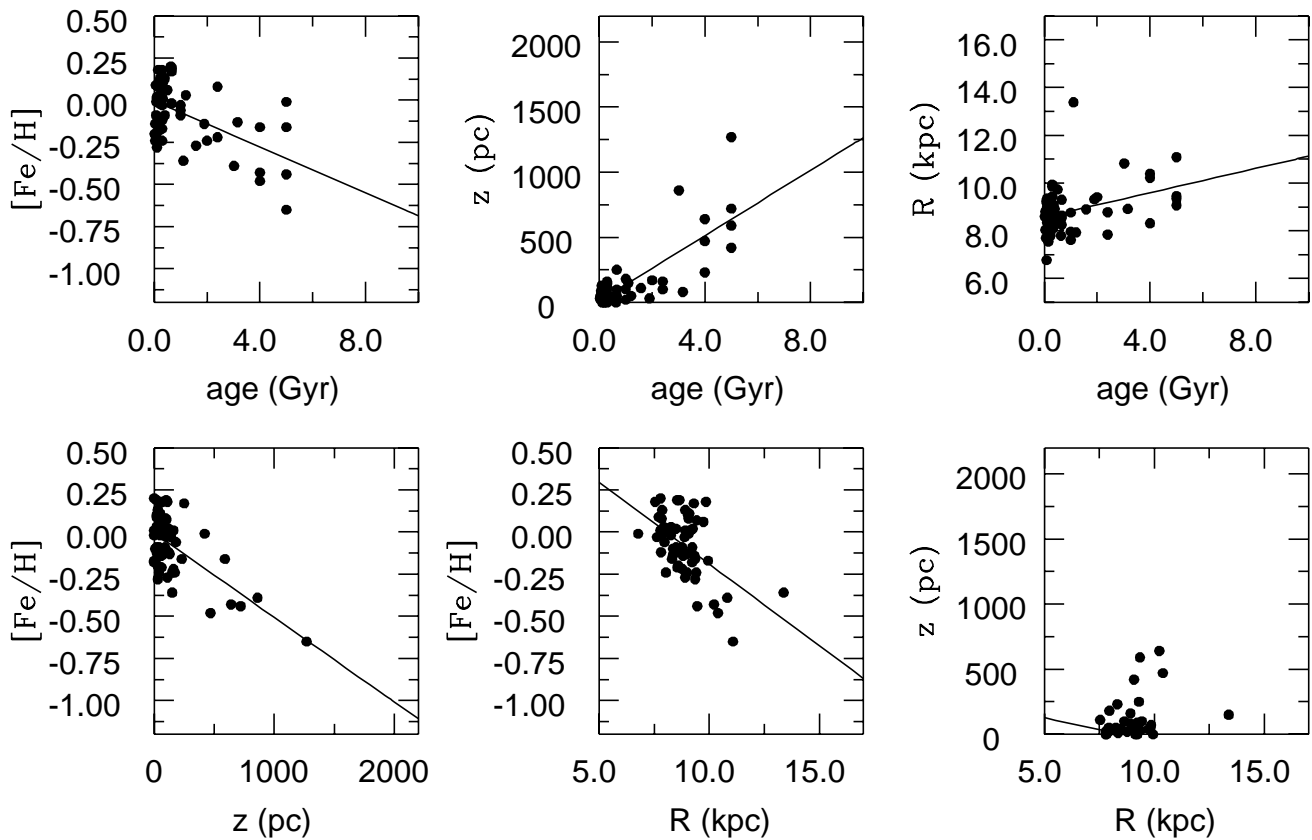,height=11.0cm,width=17.5cm}}}
\caption{Relation between the fundamental parameters of the sample
of open clusters from Piatti et~al.\ (1995),
see Sects.~2.2.2\mbox{\muspc\&\muspc}2.2.3.
Solid lines are the projections of the first principal component:
([Fe/H]\to{$t_9$}), ($z$\to{$t_9$}), (R\to{$t_9$}),
([Fe/H]\to{$z$}), ([Fe/H]\to{R}), and ($z$\to{R}).}
\label{pca_piatti}
\end{figure*}
\hfill\break
The first principal component $\psi_{1}$ gathers more than $50\%$
of the variance. Its coefficients are an indication for the degree of 
dependence it has to the original parameters. 
It appears from the first component that the clusters occupy
a three-dimensional sub-space inside the four-dimensional parameters
space: a sort of strongly elongated cigar whose main axis is almost parallel
to the age axis.
Therefore, the old open clusters in Table~1
form a one-parameter family.
\hfill\break
The projections of the first principal component on the six planes are:
([Fe/H]\to$t_{9}$), ($z$\to$t_{9}$), (R\to$t_{9}$),
([Fe/H]\to$z$), ([Fe/H]\to{R}), and ($z$\to{R}). 
They correspond approximately
to the usual regression fits between these pairs of variables.
The results are shown in Fig.~\ref{pca}.
\par
A similar analysis is made for the P95ea
data set. 
To be representative, a clusters sample has to be well 
distributed in age, metallicity and position.
Their sample includes 
62 clusters with homogeneous metallicities,
derived from DDO photometry. The other parameters are non homogeneous,
because only one metal-poor cluster (NGC 2243) with 
\mbox{$z$\muspc$>$\muspc1~kpc} above the galactic plane is included and
in addition they did not consider
NGC~6791, an old, metal-rich cluster with \mbox{$z$\muspc$=$\muspc1~kpc}. 
Moreover, about half of the sample (37 clusters) is 
younger than 1~Gyr and located at \mbox{$z$\muspc$=$\muspc100~pc}. 
Therefore, homogeneity in age and position is not 
necessarily guaranteed in their cluster sample.
In the PCA analysis we follow for consistency P95ea and
do not consider NGC 6791 and Lo 807.
The latter cluster is excluded, because it lacks an age estimate.
Concerning Ru 46 one has to keep in mind that
Carraro\mbox{\muspc\&\muspc}Patat (1995) demonstrated  
that this is likely not an open cluster.
Finally, the orbits calculated by P95ea
are based on an out-of-date Galaxy model 
(see for instance Allen\mbox{\muspc\&\muspc}Santillan 1993)
and in addition they did not correct the 
velocity components for Galactic
differential rotation (Carraro\mbox{\muspc\&\muspc}Chiosi 1994b).
\hfill\par
The results ($\psi_{1p},\psi_{2p}$) 
of the PCA are shown in Table~2, and the projection of the first
component shown in Fig.~\ref{pca_piatti}
are in the same six planes as in Fig.~\ref{pca}.
The lower eigenvalue and variance of the first principal 
component for the P95ea sample with respect to our sample
is an indication that their dataset is less homogeneous.

\subsection{Stellar population synthesis}
\label{sps}
The stellar population synthesis technique is used to generate
synthetic Hertzsprung-Russell diagrams (HRDs)
from either stellar evolutionary tracks or isochrones.
It is a powerful tool to study 
resolved stellar populations.
The evolutionary phases of a star are linked 
to each other through libraries with stellar evolutionary tracks.  
The ratio of the number of stars between different phases 
is directly related with the relative evolutionary time scale. 
This technique has been applied mainly to the analysis
of stellar aggregates (Aparicio et~al.\ 1990, Carraro et~al.\ 1993,
Aparicio \& Gallart 1995, Tosi et~al.\ 1991, Vallenari et~al.\ 1992). 
\par
\begin{figure}
\centerline{\quad\vbox{\psfig{file=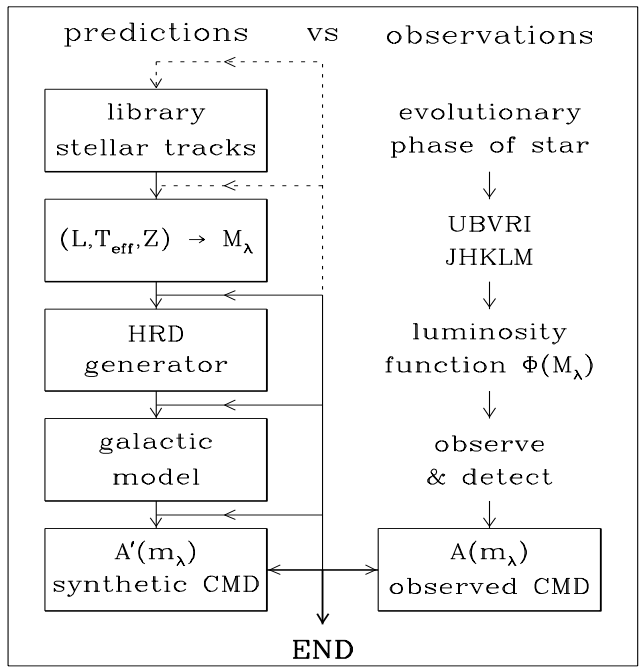,height=8.8cm,width=7.8cm}}}
\caption{Schematic diagram of the HRD-GST. Input for the 
stellar population synthesis engine is the Padova library of
stellar evolutionary tracks. 
The luminosities and effective temperature for each
synthetic star of arbitrary metallicity 
is then transformed to an absolute
magnitude in a photometric passband with the method outlined by
Bressan et~al.\ (1994) and Charlot et~al.\ (1996).
A synthetic HRD is generated,
after specification of the stellar luminosity function 
through 
the initial mass function, the star formation rate and the
age \& metallicity range.
Synthetic stars from those diagrams are then `observed'
and `detected' with the galactic model, through a Monte-Carlo technique.
In this model the density distribution of each galactic component along the 
line of sight is specified.
This results in a synthetic CMD of the field of interest.
The synthetic CMD ought to be 
comparable with the observed CMD,
when a realistic set of input parameters is used.
If there is a marginal agreement
then check the input for each step of the HRD-GST.}
\label{hrd-gst-figure}
\end{figure}

The so-called HRD galactic software telescope (HRD-GST)
is developed to study the stellar populations in our Galaxy
(Ng 1994, Ng et~al.\ 1995). The basis is formed by the latest
evolutionary tracks calculated by the Padova group
(Bertelli et~al.\ 1994 and references cited therein). 
A smooth metallicity coverage is obtained through interpolation
between the sets of tracks from low \mbox{(Z\muspc=\muspc0.0004)}
to high \mbox{(Z\muspc=\muspc0.10)} metallicity. 
Figure~\ref{hrd-gst-figure} shows a schematic diagram
of the HRD-GST, see also Ng (1994) and Ng et~al.\ (1995)
for additional details. 
\par
Synthetic CMD  are generated
with a galactic model and ought 
to be comparable with those obtained from observations.
The distribution of the stars along the line of sight is a complex 
mixture of populations. The ages, metallicities and spatial distributions
of the stars from different populations contain a wealth of 
information about the structure, formation and evolution of our Galaxy.
\hfill\break
The primary goal of the HRD-GST is to determine the interstellar
extinction along the line of sight and to obtain constraints on
the galactic structure and on the age{\to}metallicity of the
different stellar populations distinguished in our Galaxy.
The results obtained thus far 
have been reported in various papers 
(Bertelli et~al.\ 1995, 1996; Ng et~al.\ 1995\to1997).
The disc is described by a mixture of sub-populations,
each with its specific scale height and metallicity,
respectively increasing and decreasing with age.
The star formation of metal poor stars in the galactic disc 
commenced around 16\to13~Gyr ago (Ng 1994, Ng et~al.\ 1997).
We adopted 13~Gyr for this paper.
Figure~\ref{amr-figure} shows the AMR obtained with the
HRD-GST together with data from stars in the
Solar Neighbourhood 
and the line predicted by a chemical evolution model.
\par

\subsection{Chemical evolution models}
\label{chemmodels}
The AMR for nearby stars is a standard constraint for modelling the chemical
evolution of the Solar Neighbourhood, because it is a trace of the progressive
storage of heavy elements in the star-forming local interstellar medium.
\par
Most numerical or analytical chemical evolution models for the Solar
Neighbourhood and for the galactic disc
(Matteucci\mbox{\muspc\&\muspc}Fran\c{c}ois 1989; Tosi 1988;
Prantzos\mbox{\muspc\&\muspc}Aubert 1995; Ferrini et~al.\ 1992, 1994;
Pagel\mbox{\muspc\&\muspc}Tautvaisiene 1995; Timmes et~al.\ 1995;
Chiappini\mbox{\muspc\&\muspc}Matteucci 1997; Portinari et~al.\ 1997; and
references cited in those papers) follow the evolution of chemical
abundances for a mixture of stars and gas, which is assumed to be
chemically homogeneous in space. The disc is divided into concentric
rings, each of which is treated as a homogeneous region. 
Therefore, such models are aimed at reproducing average features.
\par
Many physical inputs of chemical models are rather poorly known and assumptions
need to be made about the star formation rate
(SFR), the initial mass function (IMF), the
infall time-scale, and so forth. These quantities are usually parameterized and
then calibrated on observational constraints. The predicted AMR for the Solar
Neighbourhood is sensitive to the adopted SFR, infall time-scale and IMF. All
chemical models are basically able to reproduce the observed average AMR (see,
for instance, Fig.~\ref{amr-figure}). As already mentioned in
Sect.~\ref{SolarN}, recent studies of the AMR indicate the presence of a
significant scatter with respect to the average trend. The scatter in
metallicity for a given age is comparable to the overall average increase of
metallicity from the early phases of disc evolution to the present time
(Edv93ea). As a result, the average AMR no longer
represents a tight constraint for chemical evolution models; the new challenge
lies now in reproducing the dispersion of the data, rather than the average
correlation. More complex chemical models are therefore required, which should
include mechanisms inducing the observed scatter. We briefly
summarize here various mechanisms that have been suggested. 
\hfill\break\noindent
a) {\it Diffusion of stellar orbits}\/ allows stars to move away from their
birth-places due to scattering by molecular clouds, by density waves in the
disc or by infalling satellite galaxies. This phenomenon is revealed by the
observed relation between the age and the velocity dispersion of disc stars
(Wielen 1977, Wielen et~al.\ 1992). Diffusion of stars from different
birth-places into nearby orbits results in a spread of metallicities with
respect to the expected ``local'' metallicity for a given age;
Fran\c{c}ois\mbox{\muspc\&\muspc}Matteucci (1993) discuss this effect in the
framework of chemical evolution models. Edv93ea derived
accurate kinematical data and explored the influence of stellar
orbits on the scatter in the AMR: the scatter remains large, 
even for sub-samples
with similar orbital properties. This might be an indication that another
mechanism is required to explain the scatter, but the argument is still
controversial (see Wielen et~al.\ 1996). 
\hfill\break\noindent
b) {\it Non-instantaneous mixing}\/ of stellar nucleosynthesis products in the
surrounding gas allows for self-enrichment in molecular clouds and for local
inhomogeneities (Malinie et~al.\ 1993). Pilyugin\mbox{\muspc\&\muspc}Edmunds
(1996) and van~den~Hoek\mbox{\muspc\&\muspc}de Jong (1997) combined
self-enrichment due to sequential star formation with episodic infall of
relatively metal-poor gas, triggering star formation on time-scales shorter
than the time mixing takes to smear out chemical inhomogeneities. Both
sequential star formation and infall are observed to take place in the Solar
Neighbourhood, but each mechanism in itself 
turns out to be insufficient to match
observational evidences; on the other hand, good agreement is found when both
processes are allowed to operate together. 
\hfill\break\noindent
c) {\it Different galactic sub-structures} (halo, thick disc, thin disc and
bulge), each with its own AMR, might overlap in the Solar Vicinity. Pardi
et~al.\ (1994) followed the parallel evolution of halo, thick disc and thin
disc with different evolutionary rates,
and the resulting mixture of stars shows a scatter in the AMR;
but it is
lower than the one observed in the Solar Neighbourhood. 
Most of all, this
explanation is not appealing, because 
only a low contribution from the thick disc
and halo population is expected in the Solar Neighbourhood
from star counts analysis. In fact, 
the expected ratio of metal-poor disc stars with respect to other
disc stars is about 0.9\% and the ratio of
\mbox{halo/disc\muspc$\simeq$\muspc1/1500} (Ng 1994, Ng et~al.\ 1997), 
resulting
in 2 metal-poor thick disc stars and no halo stars among the Edv93ea
F and G stars. The predicted
number of stars explains the deficiency of metal-poor stars in
Fig.~\ref{amr-figure}. In addition,    
the expected absence of halo stars is consistent with 
the disc kinematics of most of the F and
G stars in the sample. 
The observed scatter should therefore be
intrinsic to the disc and not related to the overlap with other galactic
components. 
\par
Therefore, new chemical models need to include the possibility of a more 
complex and composite evolution of the galactic disc than usually conceived;
this in turn adds new, uncertain parameters in the discussion of the chemical
history of the disc.
As correctly underlined by van den Hoek (1997) and
van~den~Hoek\mbox{\muspc\&\muspc}de~Jong (1997), 
new observational and theoretical constraints are needed to discern 
between the different processes.
\par
\par
Apart from the problem related with the scatter in the AMR, all chemical models
are able to reproduce the main observational constraints of the Solar
Neighbourhood, in spite of the different parameterization and assumptions. On
the contrary, differences appear when extending the study to the whole disc. In
particular, different predictions are given about the evolution of radial
abundance gradients with time, although all models are tuned to
reproduce the present radial gradients of oxygen and other elements. Tosi
(1996) directly compares models from different authors and outlines that in
some models the present negative gradient is rapidly established and then
remains rather unaltered, in some other models it reaches first a steeper value
than currently observed and flattens out afterwards, and yet in other models a
positive gradient is initially established, which then decreases and eventually
turns to negative values. In fact, the temporal behaviour of the gradient
basically depends on the competition between metal enrichment by stellar ejecta
and dilution by infalling metal-poor gas, i.e.\ on the SFR/infall ratio, at
different galacto-centric radii and at different ages. Since both the SFR and
the infall rate are poorly known due to the uncertainties in the related
physical processes, different assumptions can result in different predictions
about the evolution of the gradient, though all the models can reproduce the
present situation. 
An observational determination of the radial gradient at different galactic
ages would be a strong constraint in 
discriminating between different chemical models 
(see Sect.~\ref{radialgradient}).
\par
To explain the scatter in the local AMR and to describe the
evolution of the whole disc in a more consistent way, new codes with more
realistic input physics are required. A start is made with the
chemo-dynamical models of Steinmetz \& M\"uller (1994), Raiteri et~al.\ (1996),
Carraro et~al.\ (1997), and Samland et~al.\ (1997). Although arbitrary
assumptions are still needed about the SFR, 
such models can follow infall and gas
exchanges among different galactic regions self-consistently,
reducing the number of free parameters and giving a more physical picture of
the formation and evolution of the Galaxy. 

\section{Discussion}
We now continue with a comparison of the AMRs discussed in the previous
section. We first deal with the radial and vertical metallicity gradients
observed in the disc, then we proceed with a discussion of the various AMRs
and eventually suggest some improvements.

\subsection{Metallicity gradients}
To make a meaningful comparison between AMRs deduced from different samples,
they ought to be in or rescaled to the same frame of reference. Positional
dependency, due to locally different chemical evolution histories, should be
taken into account. But, when correcting for it, one has to consider that the
gradients in different age ranges need not be the same, therefore stars
or clusters with different ages are not necessarily expected to trace the same
metallicity gradient. 

\begin{table}
\caption{Radial gradient with different age-binning.}
\begin{center}
\begin{tabular}{lcc} \hline
\multicolumn{1}{c}{Age bin ($Gyr$)} &
\multicolumn{1}{c}{Bin population} &
\multicolumn{1}{c}{$\frac{d[Fe/H]}{dR}~~(dex/kpc)$} \\
\hline
$t\! < \!2$            &18& \minus0.064\mmuspc$\pm$\mmuspc0.013\\
$2\!\leq\! t\! < \!4$  &7&  \minus0.113\mmuspc$\pm$\mmuspc0.018\\
$4\! \leq\! t < \!6$   &6&  \minus0.118\mmuspc$\pm$\mmuspc0.026\\
$t\! \geq\! 6$         &6&  \minus0.079\mmuspc$\pm$\mmuspc0.033\\
\hline\hline
$t\! <\! 3$            &21& \minus0.068\mmuspc$\pm$\mmuspc0.012\\
$3\! \leq \!t \!<\! 6$ &10& \minus0.135\mmuspc$\pm$\mmuspc0.016\\
$t\! \geq\! 6$         &6&  \minus0.079\mmuspc$\pm$\mmuspc0.033\\
\hline\hline
$t\! < \!4$            &25& \minus0.086\mmuspc$\pm$\mmuspc0.009\\
$t\! \geq\! 4$         &12& \minus0.096\mmuspc$\pm$\mmuspc0.024\\
\hline\hline
$t\! < \! 5$           &29& \minus0.088\mmuspc$\pm$\mmuspc0.009\\
$t\! \geq\! 5$          &8& \minus0.061\mmuspc$\pm$\mmuspc0.029\\
\hline\hline
global                 &37& \minus0.085\mmuspc$\pm$\mmuspc0.008\\
\hline
\end{tabular}
\end{center}
\end{table}

\subsubsection{The radial abundance gradient}
\label{radialgradient}
The local AMR from Edv93ea, binned with respect to age and
radial distance, shows a radial dependence of [Fe/H] after a correction of the
radial distance for orbital diffusion. The radial gradient seems to be
shallower or absent for the oldest stars, but, within the uncertainties inherent
in the correction for orbital diffusion and the large scatter in the AMR,
the data might also be consistent with a gradient independent of age. 
\par
\begin{figure*}
\setbox1=\vbox{\hsize=10.2cm%
\psfig{file=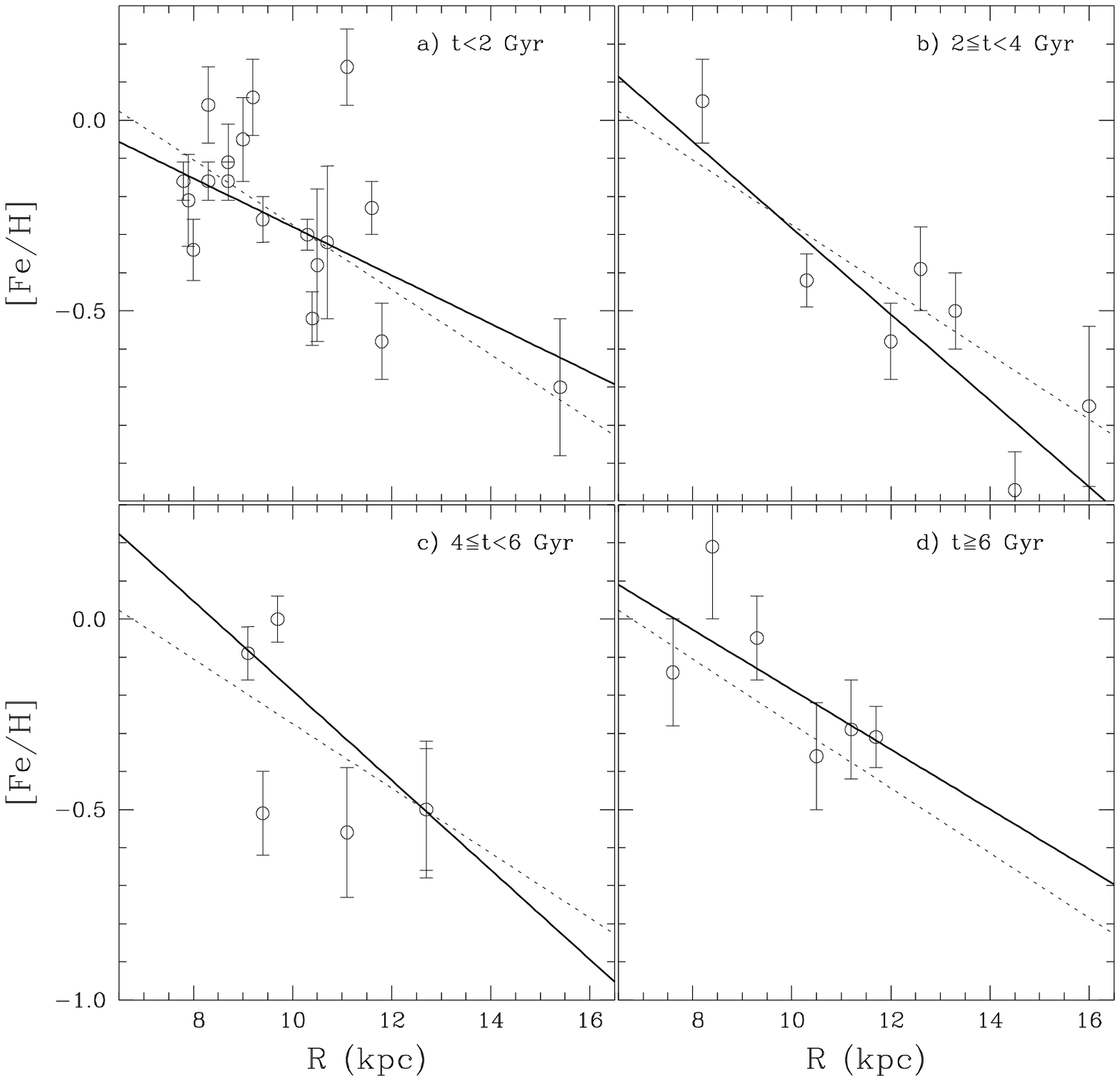,height=9.6cm,width=10.0cm}}
\setbox2=\vbox{\hsize=7.0cm%
\null\vfill\null
\caption{Radial abundance gradients for our sample
of open clusters binned in different age ranges.
The dashed line (panels {\bf a}\to{\bf e}) 
shows the average gradient (\minus0.09 dex~kpc$^{-1}$)
as determined from all the clusters from Table~1.
The solid lines in panels 
{\bf a}\to{\bf d} show the gradients determined 
for the corresponding age ranges
(\minus0.06, \minus0.11, \minus0.12, and \minus0.08  
dex~kpc$^{-1}$, respectively).
\hfill\break
\null\hfill\break
}
\label{radial}
\null\vfill\null
\centerline{\psfig{file=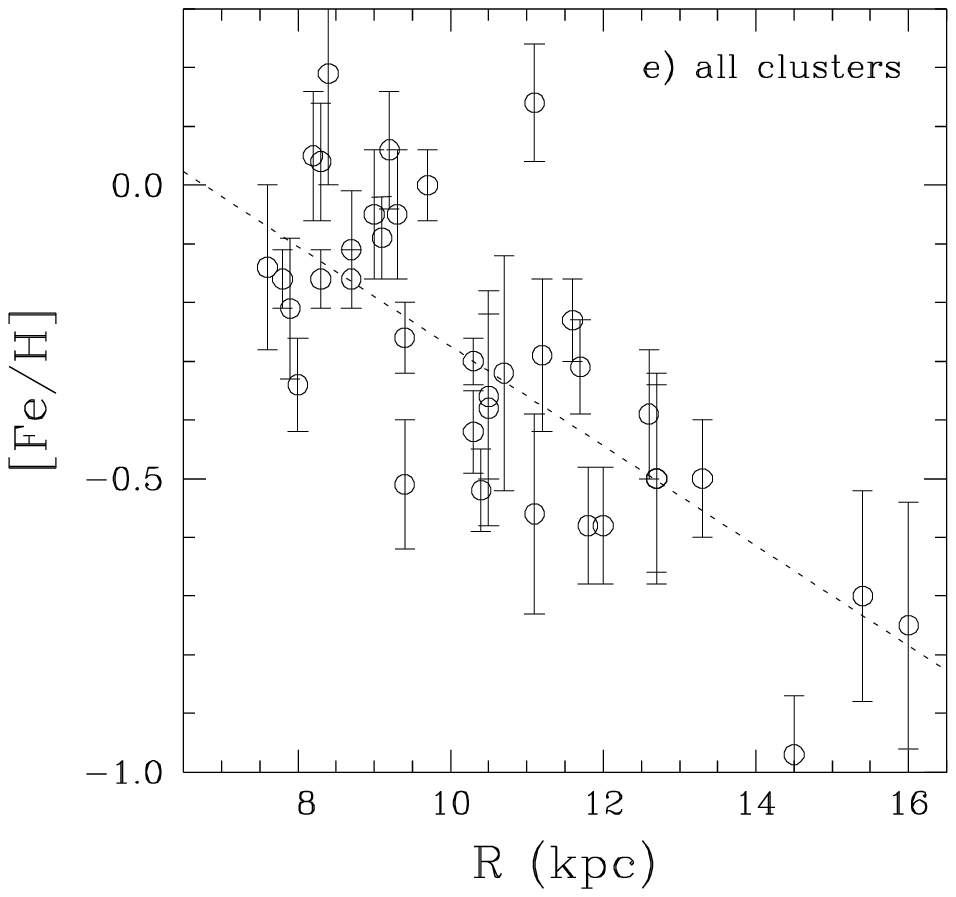,height=5.1cm,width=5.5cm}}
}
\centerline{\copy1\qquad\copy2}
\end{figure*}
A radial metallicity gradient for open clusters was first found
by Janes (1979) and then confirmed by CC94a and Friel (1995).
Open clusters are a much better tracer of the radial gradient than nearby stars,
both because they span a larger range of galacto-centric distances and because
they are not affected significantly by orbital diffusion. Indeed, orbit
calculations have shown that open clusters do not move far away from their
birth-places (Carraro\mbox{\muspc\&\muspc}Chiosi 1994b). A radial gradient is
also traced by HII-regions, B-stars, planetary nebul\ae\ (PN\ae) and so forth. 
\par
B-stars and HII-regions, due to their short lifetimes, are tracers of the
present day radial gradient (of oxygen abundance). These two different
populations seemed to lead to controversial results: the gradient traced by
\mbox{B-stars} was much more shallow (even flat) than the gradient traced by 
HII-regions (see Prantzos\mbox{\muspc\&\muspc}Aubert 1995 and 
references therein).
But the latest spectroscopic studies for large homogeneous samples of B-stars
indicate a radial gradient ${\rm d[O/H]\over{dR}}\!=\!-0.07 {\rm\ dex\
kpc^{-1}}$ (Smartt\mbox{\muspc\&\muspc}Rolleston 1997), in agreement with the
gradient deduced from HII-regions, so that nowadays the former discrepancy
appears to be solved. 
\par
As pointed out in Sect.~\ref{chemmodels}, the present radial gradient provides
a constraint for chemical evolution models, but to discern between
different models one needs to know how the gradient evolved in the past. Open
clusters are an ideal template for this analysis, since they have
well-determined ages, metallicities and galacto-centric distances.
\par
We analyse our homogeneous cluster sample of Table~1. We
divide the sample in age-bins, and for each bin we derive the gradient
by means of a weighted least-square fit in the [Fe/H] vs~R plane (using 
a version of the code by Pagel\mbox{\muspc\&\muspc}Kazlauskas 1992). In the
fitting procedure the
weights are the errors on [Fe/H], also listed in Table~1.
A similar analysis was performed by
Tosi (1995), but with a smaller sample. Our results for the four age-bins 
2 Gyr wide are shown
in Fig.~\ref{radial} and the slopes 
${\rm d[Fe/H]\over{dR}}$ of the gradient
for the different age-binning are listed in Table~3.
In panels
$b$\to$d$\/ the gradient of panel~$a$ is included for comparison. 
\par
Note
that for the determination of the radial gradient both
CC94a and P95ea did not properly
consider separate age groups. But the P95ea data set is
basically a young sample ($t\!<\!1$~Gyr) and their radial gradient of
\minus0.07~dex~kpc$^{-1}$ reflects in fact the present day gradient. 
\par
Table~3 displays the effects of different age-binning
on the derived radial gradients; indicated errors are standard deviations.
Before commenting the outcome, a caveat has to be stressed:
the number of objects involved in this analysis is relatively low, 
so statistically
the results are not very significant. They show a trend,
which requires verification with a larger sample.
\hfill\break
By selecting 2\to3~Gyr wide age-bins,
the results are similar and indicate that the present day gradient
is a bit shallower than the past one; the middle epoch seems to display a
steepening of the gradient.
The choice of wider bins (4~Gyr) basically suggests the same: the present 
day gradient appears again to be shallower. 
On the other hand, choosing a bin width of 5~Gyr, we get a different
result,
probably due to a mixing of the present and middle epoch gradients
in the younger age bin. However
in this last case the statistics is much poorer, because we are left
with only 8 clusters older than 5~Gyr versus 29 in the younger bin.
\par 
For the whole sample we derive an average gradient of
\minus0.09~dex~kpc$^{-1}$, see Fig.\ref{radial}e; 
for any age-bin in Table~3 the gradient is close to
this average value, within the errors. In the overall, our sample of clusters
seems to show that the radial gradient has not changed much in time, though a
hint can be seen for a steeper gradient in the past, at intermediate ages.
Therefore, chemical models predicting either a constant gradient or a slightly
steeper negative gradient in the past seem to be favoured with respect to
models predicting a gradient that settles to the current negative value
starting from positive initial values. For a detailed comparison of various
chemical models, we refer to Fig.~5 of Tosi (1996) and references quoted
therein. Unluckily, due to the large scatter and the small number of clusters
in the older age-bins, one cannot actually draw a firm conclusion. 
\par
The presence of a metallicity gradient for open clusters 
has been questioned by Twarog et~al.\ (1997), who find no 
gradient for the clusters in their sample at galacto-centric radius below 
10~kpc. However, their Fig.~3b displays that the majority of these clusters  
are located between \mbox{7.5\to9.5~kpc} and have an overall
spread of more than 0.2~dex in metallicity. 
Over a distance range of only 2~kpc our average gradient 
corresponds to a metallicity difference smaller than 0.2~dex. 
Therefore, within that dispersion our results are not necessarily at odds 
with their sample.
In addition, Twarog et~al.'s claim that open clusters display
a step function with a discontinuity at 10~kpc
rather than a gradient in metallicity should be
verified by separating in age groups. 
The step function could be a consequence of insufficient 
discrimination between the contribution from different age groups,
because the sample inside \mbox{$R$\muspc=\muspc10~kpc} 
is heavily weighted toward clusters younger than 1~Gyr, while the 
outer clusters are predominantly older.
\par
The radial gradient and its age dependence have been studied also by means of
PN\ae. Maciel\mbox{\muspc\&\muspc}K\"oppen (1994) find an oxygen gradient whose
present value is in agreement with that of HII-regions. It seems that the
gradient has remained roughly constant, or maybe it was slightly shallower in
the past. But, unlike open clusters, PN\ae\ have rather uncertain ages and
distances, and their present location might have been spoiled by orbital
diffusion. All these factors introduce uncertainties in the derived age
dependence of the gradient. Considering the indications from nearby stars, open
clusters and PN\ae\ altogether, the most sensible assumption is that the
gradient remained roughly constant in time. 
\par
Star counts with the HRD-GST are not sensitive enough to detect a radial
gradient within a single-age population. Indeed, a gradient of \minus0.21
dex~kpc$^{-1}$ is already induced by differences in scale-lengths among age
populations, while within a single-age population the intrinsic gradient of
\mbox{\minus0.09 dex~kpc$^{-1}$} is masked out by the adopted extinction along
the line of sight. 

\subsubsection{The vertical abundance gradient}
\label{verticalgradient}
The existence of a vertical metallicity gradient among old open clusters is
controversial. Friel (1995) and CC94a do
not find evidence for such a gradient, whereas P95ea do find a
correlation between \mbox{[Fe/H]} and vertical position.
By binning the data in $z$ they obtained a gradient equal to
\mbox{\minus0.34~dex~kpc$^{-1}$}. 
Their sample contains only one cluster above 1~kpc, 
together with 4 clusters above 500~pc. In addition they did not 
consider NGC~6791, 
which is high above the plane (1~kpc) and metal-rich. From our sample
the resulting slope is shallower: \mbox{\minus0.25~dex~kpc$^{-1}$},
not strongly dependent on any particular cluster
(Fig.~\ref{pca}, left-bottom panel).
\par
However a caveat is to be stressed, because to estimate the vertical gradient
one should disentangle the effects of vertical, radial and age
dependence. In principle, one should first bin the sample in homogeneous
sub-samples with respect to age and galacto-centric distance, and only then
analyse the vertical trend. But then, there are not enough clusters in the
bins to reliably constrain the vertical gradient. In addition, open
clusters form an incomplete sample due to tidal disruption, which is more
effective close to the galactic mid-plane. 
Therefore, clusters at low galactic latitudes
are preferentially destroyed, which might introduce a bias in the
determination of the vertical gradient. 
\par
Ng et~al.\ (1996) have demonstrated that both the radial and the vertical
distribution of open clusters contain information about the scale length and
scale height of the galactic stellar populations. In the case of field stars,
orbital diffusion is expected to be effective enough to smooth out a
vertical metallicity gradient within a single-age population, so that the
vertical structure of the disc is dominated by the different scale heights of
different age populations. Indeed, in star counts the young,
metal-richer stars are confined to regions close to the galactic mid-plane,
while the older, metal-poorer stars with a larger scale height dominate at
larger vertical distances from the galactic plane. 
As a consequence, in star counts a vertical
metallicity gradient is due to sampling of different age
populations rather than reflecting a gradient within a single-age population. 
\par
\begin{figure}
\centerline{\vbox{\psfig{file=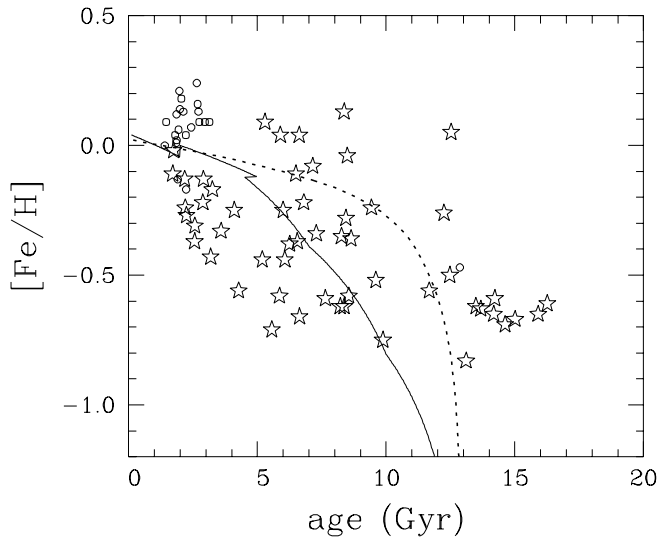,height=6.91cm,width=8.5cm}}}
\caption{The AMR for stars in the Solar Neighbourhood 
(open stars and circles, Ng\mbox{\muspc\&\muspc}Bertelli 1997),
see also Fig.~2 for details about the symbol size
and uncertainty in the age for these stars;
the solid line is the relation obtained for the galactic disc
from studies with the HRD-GST (Ng et~al.\ 1996, 1997); 
the dotted line is the prediction from the chemical evolution
model from Portinari et~al.\ (1997), with an adopted age for the disc 
of 13~Gyr.}
\label{amr-figure}
\end{figure}
In star counts, radial and vertical abundance gradients are primarily expected
due to differences between the scale length and height between different
stellar age populations. Since the relative differences in scale length 
are small
with respect to those in scale height, these latter dominate the HRD-GST star
counts analysis. A strong vertical abundance gradient in star counts is
therefore a hint that insufficient discrimination was made for the different
age groups within a data set. A metallicity gradient within an age 
population is a second order effect.
\par
Almost all open clusters of a given age are found at distances larger than
three times the exponential 
scale length and scale height for the corresponding stellar
populations of the HRD-GST. The vertical
gradient due to stellar mixes of different ages, computed at the
``transition'' region toward open clusters, is
\minus0.40 dex~kpc$^{-1}$. This value is comparable to the one found by
P95ea, confirming that their gradient could indeed be due to
insufficient discrimination of different age groups.
\par
The orbits of clusters are less affected by orbital
diffusion than those of field stars. 
So, once a vertical gradient for open
clusters were established unambiguously, it
could provide an important clue about the disc formation
history. 
But, as already stressed, current samples do not give conclusive indications.
\begin{figure}
\centerline{\vbox{\psfig{file=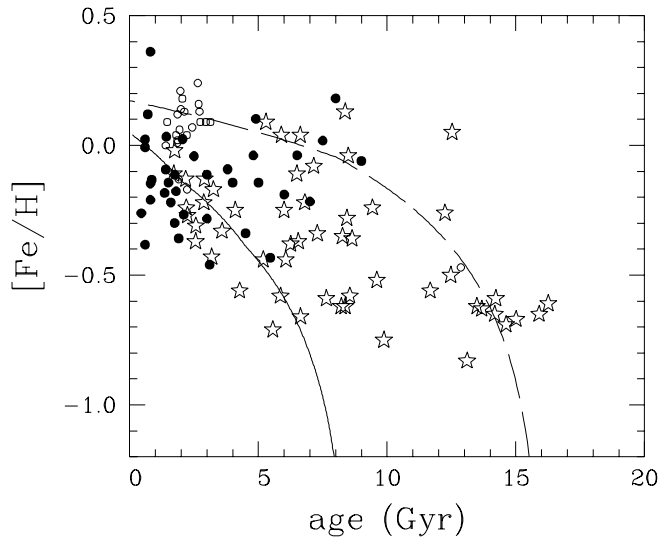,height=6.91cm,width=8.5cm}}}
\caption{The AMR for open clusters from Table~1 (filled dots)
versus the Solar Neighbourhood.
The metallicities of the open clusters are
corrected according to the global radial gradient:
[Fe/H]$_{corr}$\muspc=\muspc[Fe/H]$_{meas}-0.09$%
\muspc(R$_\odot-$\muspc{R})(kpc).
Included in this figure is a suggested improvement for the description of 
the AMR with two separate components for the HRD-GST.
The adopted age for the start of the formation of the
`young' disc component (solid line) is 8.5~Gyr and
16~Gyr for the `old' disc component (long-dashed line).
}
\label{amr-mod}
\end{figure}

\subsection{The galactic disc AMR}
Figure~\ref{amr-figure} shows various AMRs derived from:
\begin{list}{$-$}{\topsep=0pt\parsep=0pt}
\item{} {stars in the Solar Neighbourhood with the ages 
from NB98 (open circles and open stars);} 
\item{} the HRD-GST population synthesis analysis for the 
galactic disc from Ng et~al.\ (1996, 1997; solid line);
\item{} a chemical evolution model from Portinari et~al.\ (1997; dotted line).
\end{list}
The disc AMR obtained from the HRD-GST
is determined from star counts towards the north galactic pole
and the galactic centre.
No correction for radial or vertical gradients is applied, 
because star counts are not sensitive to gradients
within single-age populations, see 
Sect.~\ref{radialgradient}\muspc\&\muspc\ref{verticalgradient}.
For $t\!<\!5$~Gyr AMR(HRD-GST:disc){\muspc$\simeq$\muspc}AMR(chemics).
Between \mbox{$t$\muspc=\muspc5\to13~Gyr} the
metallicity of AMR(chemics) is higher than that of AMR(HRD-GST:disc). 
Indeed, star counts tend to follow 
the metal-poorer trend, which is more populated and
has therefore a larger weight.
\par
Using open clusters to trace the AMR has the main advantage that cluster ages
are much more reliably determined than the ages of single field stars, but also
the additional problem of the radial and/or vertical dependence of cluster
metallicity. Cameron (1985) was the first to derive an AMR from open clusters
after correcting for the radial gradient, with the aim at cleaning the data
from the space dependence. But a unique radial correction should not be applied
recklessly to different age groups, due to the possible age-dependence of the
radial gradient (Sect.~\ref{radialgradient}). Unluckily, the age dependence of
the radial gradient is poorly constrained and any confident correction for the
radial gradient of open clusters is spoiled. Our sample is consistent with a
radial gradient which is constant in time, therefore we derive from it the AMR
shown in Fig.~\ref{amr-mod} by correcting for a gradient of \minus0.09
dex~kpc$^{-1}$, independent of age. Still, one should bear in mind this
uncertainty in the correction, when comparing the AMR of open clusters with the
local AMR of nearby stars. In addition, we did not apply any correction for the
vertical gradient, since its value or even its existence are not clearly
established yet. Figure~\ref{amr-mod} displays that the AMRs of open clusters
and of stars are in good agreement, both showing a similar trend in the
scatter. We also notice that both relations show a lack of scattered points in
the metal-rich side in the age range 3\to5~Gyr and/or an excess of relatively
metal-rich objects in the range 5\to9~Gyr. 

\subsection{The role of the bulge/bar}
The up-turn of the metallicity of the open clusters
with possibly a peak near \mbox{$t\!\simeq\!8$~Gyr} might 
be related with the formation of the triaxial bar structure.
Ng et~al.\ (1996) obtained a comparable age 
and metallicity range 
(\mbox{$t$\muspc=\muspc8\to9~Gyr}, \mbox{Z\muspc=\muspc0.005\to0.030}) 
for this structure.
The bar however does not
explain the presence of metal-rich stars in the Solar Neighbourhood,
because its local density with respect to local disc stars  
is too low.
In addition some of the metal-rich stars are older than the bar.
\par
At \mbox{$t\!\simeq\!8$~Gyr}, the stars in the Solar Neighbourhood are
metal-richer than expected from the AMR(HRD-GST:disc) and AMR(chemics). The
calculations from Samland et~al.\ (1997) suggest that this could be due to
pre-enrichment with material originating from type~II SN\ae\  from the bulge.
On the other hand, if the `bar' structure formed through a merger event
some of the metal-richer stars from the inner disc or bulge regions could have
migrated near to the Solar Neighbourhood and settled down there. This could
explain the presence of rather metal-rich clusters and stars between 5 and
9~Gyr. But suitable dynamical models are
required (Mihos{\muspc\&\muspc}Hernquist 1996)
to explore and check the merger/capture scenario properly
against all the observational constraints.

\subsection{Improvements}
The comparison of the AMRs from various methods 
suggests some improvements. 
\hfill\break\noindent
a) {\it The Solar Neighbourhood}\/: 
a new sample of stars with spectroscopic metallicities
and reliable ages is desirable and
better ages are needed for $t\!>\!10$~Gyr. 
\hfill\break\noindent
b) {\it Open \& globular clusters}\/:
the sample of open clusters should be enlarged 
to verify the lack of relatively metal-rich objects 
in the age range 3\to5~Gyr, since the apparent `U-shape'
of the AMR might provide clues about infalling
and/or merger events.
\hfill\break
The old, metal-rich open clusters and the young, metal-rich globular
clusters could indicate when the transition between the two types of
clusters actually occurred and provide a clue 
if this could be initiated by the formation of the galactic bar.
The high galactic fore- and background contamination in the 
colour-magnitude diagrams of the metal-rich globular clusters suggests a 
combined approach with a galactic population model, to obtain a 
self-consistent interpretation about the age, metallicity, distance
and extinction towards the clusters.
\hfill\break\noindent
c) {\it Stellar population synthesis}\/:
a second, old and metal-rich disc component should be included 
in the HRD-GST description of the disc (see Fig.~\ref{amr-mod})
and its nature investigated.
\hfill\break\noindent
d) {\it Chemical evolution models}\/:
possible pre-enrichment and/or influence of the bulge
and of the bar should be considered to explain the metal-rich part
of the AMR at old ages.
Chemo-dynamical models look promising in this respect.
In particular, with the lagrangian approach (Raiteri et~al.\ 1996, 
Carraro et~al.\ 1997) 
the behaviour of the gas settling down in the disc,
vertical and radial infall, outflows and so on can be followed.
In addition, such models are
suitable to study merging and capture events.

\section{Conclusions}
We compared the galactic disc AMRs obtained from four different points of view.
Our results are here briefly outlined. 
\hfill\break
$\bullet$\ 
For the first time a multivariate analysis of a large, homogeneous sample
of open clusters was performed and the correlations between cluster
parameters were unravelled with an {\it a priori}\/ approach.
\hfill\break
$\bullet$\ 
We investigated the radial gradient and its time evolution, although
a larger sample of open clusters would be required to constrain the past
behaviour of the gradient better.
\hfill\break
$\bullet$\
We showed that an apparent strong vertical gradient in open cluster
is likely due to insufficient discrimination between different age groups.
\hfill\break
$\bullet$\ 
Considering our improved sample open clusters and a sub-sample of local stars
with reliable ages, both sets of objects trace a similar AMR, whose peculiar
shape in the range 3\to9 Gyr ago might give clues about past infall or merger
events. 

\section*{Acknowledgments}
G.\ Carraro thanks Prof.\ E.\ Cappellaro for kind assistance in using SPSS.
G.\ Bertelli is acknowledged for valuable suggestions and discussions. 
We also thank the anonymous referee for constructive suggestions. This
research is supported by the TMR grant ERBFMRX-CT96-0086 from the European
Community (Network: Formation and Evolution of Galaxies), by the Italian
Ministry of University, Scientific Research and Technology (MURST) and by the
Italian Space Agency (ASI). L.~Portinari also acknowledges financial support
from the Danish Rektorkollegiet.

\end{document}